\newif\ifPTEP
\newif\ifarxiv

\PTEPfalse \arxivtrue

\ifPTEP
\documentclass[preprint,pteplogo]{ptephy_v2}
\preprintnumber{2607.15335}
\fi
\ifarxiv
\documentclass[aps,prd,twocolumn,nofootinbib,superscriptaddress]{revtex4-2}
\fi
\usepackage{amsmath}
\usepackage{amssymb}
\usepackage{graphicx}
\usepackage{xcolor}
\usepackage{bm}
\usepackage{braket}
\usepackage{aas_macros}
\usepackage{hyperref}

\begin{document}

\title{Primordial tensor mode from the neutrino sector}

% PTEP %%%%%%%%%%%%%%%%%%%%%%%%%%%
\ifPTEP
\author[1,2]{Shohei~Saga}
\affil[1]{Institute for Advanced Research, Nagoya University, Furo-cho, Chikusa-ku, Nagoya 464-8601, Japan}
\affil[2]{Kobayashi-Maskawa Institute for the Origin of Particles and the Universe, Nagoya University, Nagoya, Aichi 464-8602, Japan}

\author[2,3,4]{Shuichiro~Yokoyama}
\affil[3]{Department of Physics, Nagoya University, Nagoya, Aichi 464-8602, Japan}
\affil[4]{Kavli Institute for the Physics and Mathematics of the Universe (WPI), The University of Tokyo, Kashiwa, Chiba 277-8583, Japan}
\subjectindex{E60, E63, E65}
\fi
% PTEP %%%%%%%%%%%%%%%%%%%%%%%%%%%

% arxiv %%%%%%%%%%%%%%%%%%%%%%%%%%%
\ifarxiv
\author{Shohei Saga}
\affiliation{Institute for Advanced Research, Nagoya University, Furo-cho, Chikusa-ku, Nagoya 464-8601, Japan}
\affiliation{Kobayashi-Maskawa Institute for the Origin of Particles and the Universe, Nagoya University, Nagoya, Aichi 464-8602, Japan}

\author{Shuichiro~Yokoyama}
\affiliation{Kobayashi-Maskawa Institute for the Origin of Particles and the Universe, Nagoya University, Nagoya, Aichi 464-8602, Japan}
\affiliation{Department of Physics, Nagoya University, Nagoya, Aichi 464-8602, Japan}
\affiliation{Kavli Institute for the Physics and Mathematics of the Universe (WPI), The University of Tokyo, Kashiwa, Chiba 277-8583, Japan}
\fi
% arxiv %%%%%%%%%%%%%%%%%%%%%%%%%%%

\begin{abstract}%
We show that a regular solution for primordial tensor fluctuations can arise from the higher-order multipoles of the collisionless neutrino distribution after neutrino decoupling. Focusing on the leading case with the neutrino octupole mode (tensor-octupole mode), we derive the initial conditions for the Einstein-Boltzmann system and calculate angular power spectra for the cosmic microwave background (CMB) anisotropies. Compared with the standard gravitational-wave mode, the tensor-octupole mode has a weaker large-scale metric response. It therefore gives a suppressed reionization bump and a different oscillation phase in the tensor CMB power spectra, providing a way to separate it from the standard tensor mode.
\end{abstract}

\maketitle

\section{Introduction}
\label{sec: introduction}

Primordial tensor perturbations provide a unique probe of the very early universe. They can be generated by quantum fluctuations during inflation, or by other high-energy processes in the early universe (e.g., Refs.~\cite{1979JETPL..30..682S,1982PhLB..115..189R,1982PhRvL..49.1110G,1981JETPL..33..532M}). One of their most characteristic probes is the $B$-mode polarization of the cosmic microwave background (CMB) anisotropies~\cite{1997PhRvD..55.1830Z,1997PhRvD..55.7368K,1997PhRvD..56..596H}. A detection of $B$ modes induced by the primordial tensor mode would provide direct information on physics at energy scales much higher than those accessible by terrestrial experiments.

In most CMB analyses, primordial tensor perturbations are assumed to start from the standard regular GW initial condition. In this mode, the tensor metric perturbation is finite and nonzero on super-horizon scales during the radiation-dominated era.
We refer to this standard initial condition for the primordial tensor perturbations as the \textit{GW mode}. The power spectrum of the primordial tensor perturbations is conventionally characterized by the tensor-to-scalar ratio $r$ and the tensor spectral index $n_t$. The current observations are consistent with a vanishing GW mode and provide stringent upper bounds on $r$~\cite{2020A&A...641A...6P,2021PhRvL.127o1301A}, and the future CMB experiments are expected to improve the sensitivity to $r$ by an order of magnitude~\cite{2020SPIE11443E..2FH,2023PTEP.2023d2F01L}.

Although this GW mode is well motivated by the inflationary tensor perturbations, it is not the only possible regular initial condition in a multi-component Einstein-Boltzmann system. A similar situation has been well known in the scalar sector. In addition to the adiabatic mode, which is the simplest and observationally preferred initial condition, several regular isocurvature modes can be constructed. Such modes arise naturally in multi-field inflationary models (see, e.g., Ref.~\cite{2009PhR...475....1M} for a review).
In these modes, the primordial perturbations are stored in the relative density or velocity perturbations among different species, rather than only in the total curvature perturbation. Although scalar isocurvature modes are already tightly constrained by CMB observations, they remain useful probes of physics beyond the minimal adiabatic scenario (see Ref.~\cite{2014A&A...571A..22P} and references therein).

The vector sector also gives a related example. Vector metric perturbations decay in a perfect-fluid universe without anisotropic stress, and hence they are often neglected. However, free-streaming species can support regular vector perturbations through their anisotropic stress \cite{1992ApJ...392..385R,1994PhRvD..50.2541R,2004PhRvD..70d3518L}. 
Such vector modes also generate CMB anisotropies, including the $B$-mode polarization, and can be distinguished from the standard tensor contribution by their different angular spectra \cite{2012PhRvD..85d3009I,2014JCAP...10..004S}.
More recently, vector initial conditions in which the primordial information is carried by the neutrino octupole, as well as vector modes sourced by an early anisotropic stress, have been considered \cite{2025JCAP...10..112K,2026arXiv260508907R,2026arXiv260513016Y}. Although realizing these initial conditions requires nontrivial early-universe physics, they are mathematically allowed in the Einstein-Boltzmann system and provide a useful phenomenological extension of the standard cosmological model.

Motivated by these studies, we ask whether the tensor sector can also have a regular initial condition in which the primordial information is stored in the neutrino phase-space distribution rather than in the standard GW metric amplitude.
Such a mode can be constructed in the Einstein-Boltzmann hierarchy; in fact, it is contained in the general regular solutions for tensor perturbations in the presence of collisionless matter obtained by Rebhan and Schwarz \cite{1994PhRvD..50.2541R}, as we show explicitly in Sec.~\ref{sec: initial_conditions}. Our aim is to isolate it as an independent primordial initial condition, in analogy with the isocurvature modes in the scalar sector, and to study its observational consequences. In this paper, we focus on the leading realization sourced by the neutrino octupole and investigate its CMB signatures by modifying the public Boltzmann code. In this case, the octupole first generates the quadrupole, and then the quadrupole acts as the tensor anisotropic stress and induces the metric perturbation, i.e., gravitational wave background. We call this solution the tensor-octupole (OCT) mode. Accordingly, GWs are generated dynamically and are initially zero. This leads to CMB spectra that differ from those of the standard GW mode.

This paper is organized as follows. In Sec.~\ref{sec: basic}, we introduce the tensor Einstein-Boltzmann equations. In Sec.~\ref{sec: initial_conditions}, we discuss the regularity condition, and derive the tensor-OCT initial conditions. In Sec.~\ref{sec: numerical_results}, we present the numerical results by modifying the public Boltzmann code, and provide the parameter constraints based on the current CMB data. Section~\ref{sec: summary} is devoted to the discussions and the summary of this paper.

\section{Basic equations}
\label{sec: basic}

We start from a spatially flat Friedmann-Lemaitre-Robertson-Walker metric with a tensor perturbation
\begin{align}
{\rm d}s^2 = a^2(\eta)\left[-{\rm d}\eta^2+\left(\delta_{ij}+h_{ij}\right){\rm d}x^i{\rm d}x^j\right],
\end{align}
with $\eta$ and $a(\eta)$ being the conformal time and the scale factor, respectively.
We decompose the tensor metric perturbation into the two helicity states as
\begin{align}
h_{ij}(\eta,\bm{x})=\int\frac{{\rm d}^{3}\bm{k}}{(2\pi)^{3}}\sum_{\lambda=\pm2}h_\lambda(\eta,\bm{k})e_{ij}^{(\lambda)}(\hat{\bm{k}}) e^{i\bm{k}\cdot\bm{x}},
\end{align}
with $e_{ij}^{(\lambda)}(\hat{\bm{k}})$ being the polarization tensor, obeying the transverse and traceless conditions, $k^ie_{ij}^{(\lambda)}=0$ and $e_{ii}^{(\lambda)}=0$. Hereafter, we omit the helicity label as we assume the same power for the two helicities and hence parity invariance.

The Einstein equation of the tensor mode with the neutrino anisotropic stress is given by
\begin{align}
\ddot{h} + 2\mathcal H \dot{h} + k^2 h = \mathcal H^2 R_\nu \pi_{\nu,2},
\label{eq: tensor_einstein}
\end{align}
where a dot denotes a derivative with respect to the conformal time $\eta$, and $\mathcal H=\dot{a}/a$. We also define $R_\nu := \bar{\rho}_{\nu}/(\bar{\rho}_{\nu}+\bar{\rho}_{\gamma})$ with $\bar{\rho}_{\nu}$ and $\bar{\rho}_{\gamma}$ being the background energy densities of neutrinos and photons, respectively. 
Following the notation of Ref.~\cite{2010PhRvD..81d3517S}, we relate the neutrino anisotropic stress to the tensor brightness quadrupole by $\pi_{\nu,2}=2I_{\nu,2}/5$. Only the quadrupole directly contributes to the stress-energy tensor in Eq.~\eqref{eq: tensor_einstein}. Here, we assume that the neutrinos are massless for simplicity.
The collisionless tensor neutrino hierarchy is
\begin{align}
\dot{I}_{\nu,2} + \frac{\sqrt5}{7}k I_{\nu,3} + 4\dot{h}=0, \label{eq: I2eq}
\end{align}
for the quadrupole and, for $\ell\ge3$,
\begin{align}
\dot{I}_{\nu,\ell} 
+ k\left[ \frac{\sqrt{(\ell+1)^2-4}}{2\ell+3}I_{\nu,\ell+1} - \frac{\sqrt{\ell^2-4}}{2\ell-1}I_{\nu,\ell-1} \right]=0 . \label{eq: Ielleq}
\end{align}
The moments with $\ell\ge3$ describe independent phase-space degrees of freedom, but they do not enter the Einstein equation directly. They affect the metric only after their amplitude is transferred to the quadrupole through the hierarchy. This property is important for the regularity argument given below. We now use these equations to identify the lowest regular tensor initial condition sourced by the neutrino hierarchy.

\section{Initial conditions}
\label{sec: initial_conditions}

We now derive the regular tensor initial condition with a vanishing initial metric amplitude and a nonzero primordial neutrino octupole. We first explain why the octupole is the leading neutrino moment to consider. The quadrupole, or the anisotropic stress, enters Eq.~(\ref{eq: tensor_einstein}). If a nonzero constant quadrupole is imposed without any compensating source, Eqs.~(\ref{eq: tensor_einstein}) and (\ref{eq: I2eq}) imply a logarithmically singular metric perturbation and also octupole moment in the limit $k\eta\to0$. 
Such initial conditions do not realize a regular super-horizon solution
of the collisionless Einstein--Boltzmann system. A constant anisotropic stress can be allowed when another component compensates the total anisotropic stress~\cite{2004PhRvD..70d3011L,2010PhRvD..81d3517S}. Another possibility is that the anisotropic stress is generated at a certain finite time and present for a finite interval. In this case, the metric perturbation acquires a logarithmically enhanced regular mode, the so-called passive mode \cite{2004PhRvD..70d3011L,2010PhRvD..81d3517S}. This passive component however evolves with the same transfer function as the standard GW mode, and does not provide a qualitatively new tensor mode. We do not include such a passive contribution, but focus on the more nontrivial initial condition.

As shown in Eq.~\eqref{eq: tensor_einstein}, the octupole does not directly enter the Einstein equation. Instead, a constant $I_{\nu,3}$ generates the quadrupole through Eq.~(\ref{eq: I2eq}), leading to $I_{\nu,2}=\mathcal{O}(x)$. The resulting quadrupole then induces $h=\mathcal{O}(x)$, so that both the quadrupole and the tensor metric perturbation smoothly vanish as $x\rightarrow0$. We call this solution the tensor-OCT mode. The same argument can be applied when the first nonzero moment is $I_{\nu,L}$ with $L\ge3$. The downward propagation through the collisionless hierarchy gives $I_{\nu,\ell}=\mathcal{O}\left(x^{L-\ell}\right)$ for $2\le\ell<L$, and hence $I_{\nu,2}=h=\mathcal{O}\left(x^{L-2}\right)$. Among this family, the octupole with $L=3$ is the leading mode and is expected to produce the largest metric and CMB responses for the same primordial normalization. We note that, as the standard GW mode is also regular, taking $x\to 0$ leads to the tensor metric perturbation being a non-vanishing constant while the neutrino multipoles vanish.

Expanding the scale factor in the radiation-dominated era, 
\begin{align}
a(\eta) \simeq a_0\frac{\Omega_r}{\Omega_m} \left(k_{\rm eq}\eta+\frac14k_{\rm eq}^2\eta^2\right) , \label{eq: scale_factor}
\end{align}
where $\Omega_r$ and $\Omega_m$ are the present radiation and matter density parameters, $a_0$ is the present scale factor, and $k_{\rm eq}$ is the wavenumber of the mode that enters the horizon at matter-radiation equality, $k_{\rm eq}=a_{\rm eq} H_{\rm eq} = \Omega_m H_0/\sqrt{\Omega_r}$, with $H_0$ being the present Hubble parameter.
The Einstein-Boltzmann system (\ref{eq: tensor_einstein})--(\ref{eq: Ielleq}) can then be solved order by order in $x=k\eta$, and keeping the leading correction proportional to $k_{\rm eq}/k$, we obtain
\begin{align}
h & \simeq -\frac{\sqrt5 R_\nu}{7(4R_\nu+5)}x -\frac{5\sqrt5R_\nu}{28(4R_\nu+5)(4R_\nu+15)}\frac{k_{\rm eq}}{k}x^2,
\label{eq: oct_h}
\\
I_{\nu,2} &\simeq -\frac{5\sqrt5}{7(4R_\nu+5)}x + \frac{5\sqrt5R_\nu}{7(4R_\nu+5)(4R_\nu+15)}\frac{k_{\rm eq}}{k}x^2,
\label{eq: oct_I2}
\\
I_{\nu,3} &\simeq 1-\frac{16R_\nu+35}{42(4R_\nu+5)}x^2,
\\
I_{\nu,4} &\simeq \frac{2\sqrt3}{7}x, \quad
I_{\nu,5} \simeq \frac{\sqrt{7}}{21}x^2,
\end{align}
where we set the initial octupole amplitude to unity. For the standard radiation density, the leading coefficient of the metric perturbation in Eq.~(\ref{eq: oct_h}) is only about $0.02$. Hence, even around horizon crossing, the metric perturbation generated from a unit octupole is strongly suppressed. This weak conversion from the octupole to the metric perturbation is important for the CMB signatures. 

We comment on the relevance of the present tensor-OCT solution to the previous works. The tensor-OCT solution is a particular member of the general regular tensor solutions of Ref.~\cite{1994PhRvD..50.2541R}. More specifically, in their notation, $H(0) = 0$, $\beta^{(2)}_{1} \neq 0$, and $\beta^{(2)}_{n\neq1} = 0$ correspond to the tensor-OCT mode. This satisfies their regularity condition $\beta^{(2)}_{0} + 2\beta^{(2)}_{2}/7 + \beta^{(2)}_{4}/21 = 0$, and selects the super-horizon growing solution. 
In Ref.~\cite{2026JHEP...06..216O}, they recently study the universal collisionless response in a real-time thermal field-theory calculation. The tensor-OCT mode can be regarded as a particular choice of homogeneous matter initial data in this transport description (see Appendix B of Ref.~\cite{2026JHEP...06..216O}).
The present analysis specializes this existing solution to a neutrino octupole and develops its CMB transfer functions, observational signatures, and observational constraints.

\begin{figure}[t]
\centering
\includegraphics[width=0.49\textwidth]{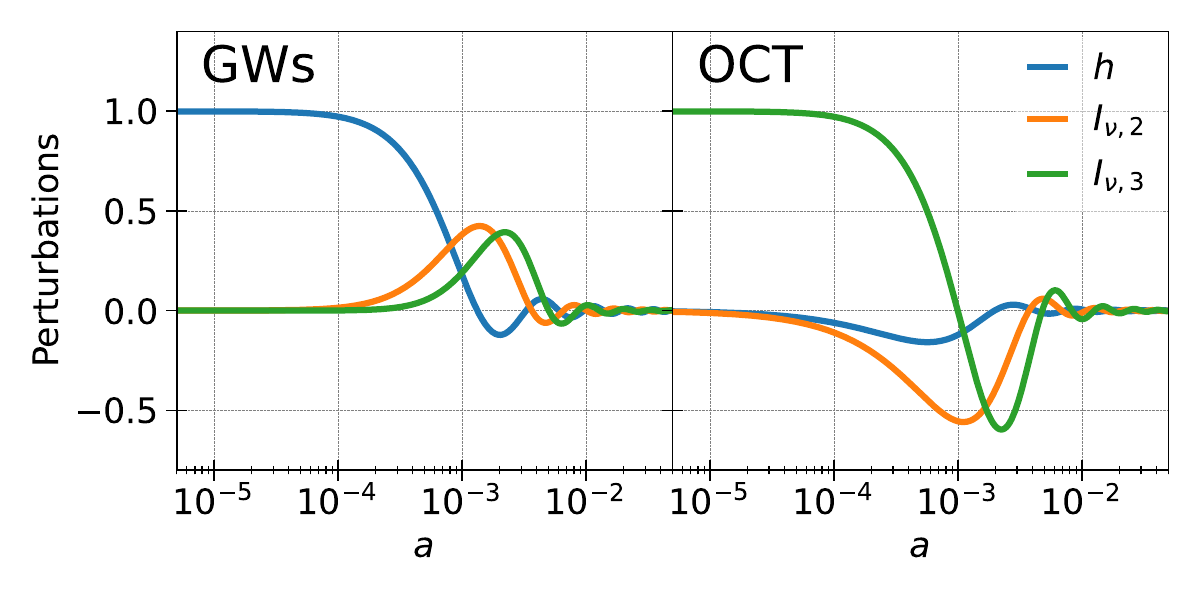}
\caption{Time evolution of the perturbations $h$, $I_{\nu,2}$, and $I_{\nu,3}$ for the standard GW initial condition (left) and the tensor-OCT initial condition (right), at the wavenumber $k=0.01\, {\rm Mpc}^{-1}$.}
\label{fig:transfer}
\end{figure}

Figure~\ref{fig:transfer} compares the time evolution of the perturbations for the standard GW mode and the tensor-OCT mode. For the GW mode, the metric perturbation initially stays constant, and the neutrino free streaming generates $I_{\nu,2}$ and $I_{\nu,3}$. The quadrupole then affects the metric perturbation through the anisotropic stress. On the other hand, for the tensor-OCT mode, the order is opposite: $I_{\nu,3}$ is primordial and initially constant, and $I_{\nu,2}$ and $h$ are generated during the evolution. On super-horizon scales, $I_{\nu,3}$ remains constant, while $I_{\nu,2}$ and $h$ grow. These behaviors agree with Eqs.~(\ref{eq: oct_h}) and (\ref{eq: oct_I2}). Around horizon crossing, the octupole transfers its amplitude both to the quadrupole and to the higher moments. The induced metric perturbation then starts to oscillate with a phase different from that of the GW mode.

\section{Numerical results}
\label{sec: numerical_results}

In this section, we present the numerical results of the tensor-OCT mode. We first describe the setup of the primordial statistics, and then show the numerical results of the CMB signals.

\subsection{Setup of primordial statistics}

We consider a primordial tensor mode with a vanishing initial metric perturbation and a nonzero octupole, as discussed in Sec.~\ref{sec: initial_conditions}. Let $X(\bm{k})$ represent either the initial GW amplitude $h$ or the initial neutrino octupole amplitude $I_{\nu,3}$. 
We assume that the primordial OCT amplitude is a statistically homogeneous and isotropic Gaussian random field, and define their dimensionless power spectra as
\begin{align}
\Braket{X(\bm{k})X^{*}(\bm{k}')}
 = (2\pi)^{3}\delta^{(3)}(\bm{k}-\bm{k}')
 \frac{2\pi^2}{k^3} \mathcal{P}_{X}(k) .
 \label{eq: two_point}
\end{align}
We parametrize the primordial spectrum for the standard GW initial condition and the tensor-OCT initial condition, respectively, as
\begin{align}
\mathcal{P}_{h}(k)=A_{\rm s} r_{t}\left(\frac{k}{k_{*}}\right)^{n_{t}}, \quad 
\mathcal{P}_{I_{\nu,3}}(k)=A_{\rm s} r_{\nu} \left(\frac{k}{k_{*}}\right)^{n_{\nu}}.
\end{align}
Here, the parameters $A_{\rm s}$, $r_{t}$, $n_{t}$, $r_{\nu}$, and $n_{\nu}$ are the scalar amplitude, tensor-to-scalar ratio for the GW mode, spectral index for the GW mode, tensor-to-scalar ratio for the OCT mode, and spectral index for the OCT mode, defined at the pivot scale $k_{*}=0.05\,{\rm Mpc}^{-1}$. Strictly speaking, the tensor-to-scalar ratio for the OCT mode $r_{\nu}$ is the amplitude of the primordial octupole normalized by the scalar amplitude, and is not the usual tensor-to-scalar ratio describing the metric perturbation. The two ratios $r_{t}$ and $r_{\nu}$ are not directly comparable, since they describe different physical quantities. However, we will use the same notation for both ratios for convenience.

Although we do not specify a microscopic generation model for the tensor-OCT mode, we can discuss its infrared behavior under an analyticity assumption motivated by causality and finite-range correlations (see, e.g., Ref.~\cite{2003JCAP...11..010D}). If the dimensional octupole spectrum $2\pi^2\mathcal{P}_{I_{\nu,3}}(k)/k^3$ is analytic at $\bm{k}=\bm{0}$, statistical isotropy implies an expansion in even powers of $k$.
For a nonzero constant (white-noise) term, the dimensionless spectrum therefore behaves as
\begin{align}
\mathcal{P}_{I_{\nu,3}}(k)
\propto k^{3}
\left(1+\sum_{n\geq1}c_{2n}k^{2n}\right),
\end{align}
with $c_{2n}$ being constants. Thus, a nonzero white-noise term gives the dimensionless infrared index $n_{\nu} = 3$ and additional constraints that eliminate this term can make the spectrum steeper. A red or nearly scale-invariant spectrum in this infrared regime requires pre-existing long-range correlations or a generation history outside the finite-range assumptions above. For example, correlations inherited from inflation would have to survive in a parent degree of freedom and be transferred to the neutrino distribution after its collisional damping becomes ineffective. 
The tensor-OCT initial condition can be applied to massless free-streaming dark radiation, which indeed obeys the same transport hierarchy, and can be an interesting possibility for more realistic generation models. However, these model-dependent discussions are beyond the scope of this paper, and we will treat $n_{\nu}$ as a phenomenological free parameter in the following analysis.

\subsection{Numerical results of the tensor-OCT mode}

We implemented the tensor-OCT mode as an additional tensor initial condition in a modified version of CAMB \cite{2000ApJ...538..473L,2011ascl.soft02026L}. This section presents the numerical results of the transfer functions and CMB angular power spectra. We also perform a Markov-chain Monte Carlo (MCMC) analysis to constrain the amplitude and spectral index of the OCT mode using recent observational data.

With the definitions of the power spectra~(\ref{eq: two_point}), we calculate the CMB angular power spectra for the GW and OCT modes, explicitly given by
\begin{align}
C_{\ell}^{AA'}=\int{\rm d}\ln k\, \mathcal{P}_{X}(k)\Delta_{\ell}^{A}\Delta_{\ell}^{A'},
\end{align}
where we define the transfer function $\Delta_{\ell}^{A}$ for the CMB anisotropies of $A,\, A' = T,\, E,\, B$, which are calculated by solving the Einstein-Boltzmann system. As long as parity is conserved, the non-vanishing angular spectra are $TT$, $EE$, $TE$, and $BB$. In this work, we assume that the GW and OCT initial amplitudes are statistically uncorrelated.

\begin{figure*}[t]
\centering
\includegraphics[width=\textwidth]{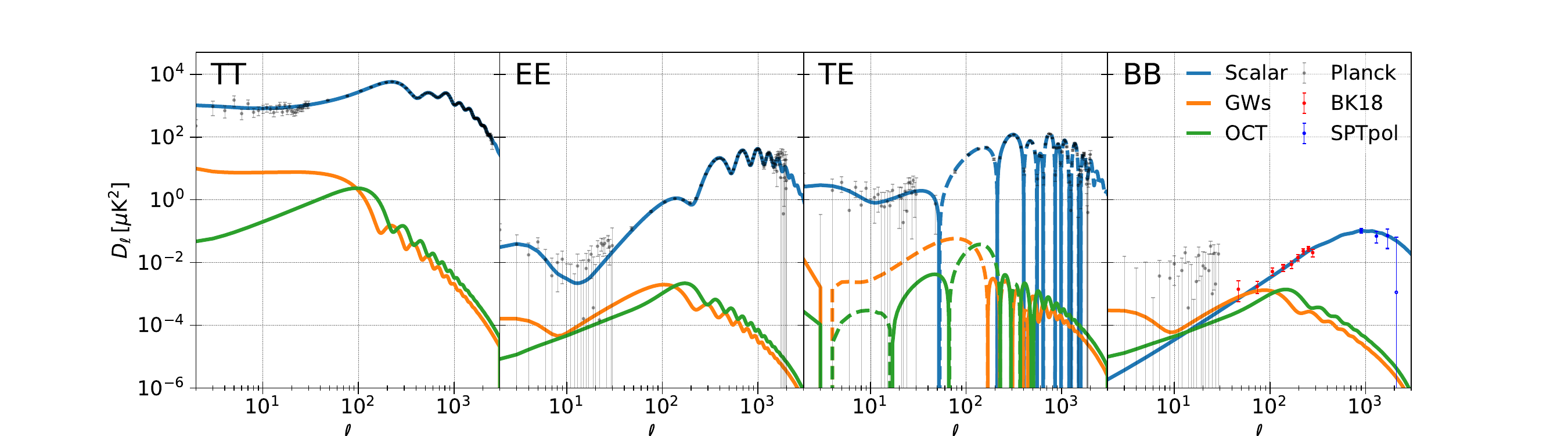}
\caption{CMB angular power spectra for the standard GW mode (orange) and the tensor-OCT mode (green), compared with the adiabatic scalar contribution (blue). We also show the observed power spectra from Planck~\cite{2020A&A...641A...5P}, BICEP/Keck 2018 (BK18)~\cite{2021PhRvL.127o1301A}, and SPTpol~\cite{2020PhRvD.101l2003S,2025OJAp....847459B} data. Negative SPTpol bandpowers are shown at their absolute values using open markers. We set the tensor-to-scalar ratio to $r_{t}=0.02$ and $r_{\nu}=15$ for the GW and OCT modes, respectively, and the spectral indices to $n_{t}=n_{\nu}=0$, for illustration purposes. The standard cosmological parameters are fixed to the Planck 2018 best-fit values~\cite{2020A&A...641A...6P}. For the $TE$ spectrum, we show negative values by dashed lines.}
\label{fig:spectra}
\end{figure*}

Figure~\ref{fig:spectra} shows representative CMB angular power spectra, where $D^{AA'}_\ell \equiv \ell(\ell+1)C^{AA'}_\ell/(2\pi)$. 
Looking at the CMB spectra at large scales, the OCT mode is suppressed compared to the standard GW mode. The GW mode has a constant metric perturbation before horizon crossing. In contrast, the OCT mode does not have a constant metric component and instead generates only the suppressed metric perturbation given in Eq.~(\ref{eq: oct_h}), i.e., $h \propto k\eta$. As a result, the OCT mode less efficiently produces the CMB anisotropies at the larger angular scales.
The same difference is also clearly seen for the suppressed reionization feature in the polarization spectra. We note that the suppression of the reionization bump is also seen in the vector OCT mode \cite{2025JCAP...10..112K,2026arXiv260508907R,2026arXiv260513016Y}. Schematically, the reionization bump is generated by the line-of-sight integral of the photon quadrupole at reionization, which is sourced by the tensor metric perturbation. 
Because the metric perturbation is generated differently in the GW and OCT modes, the photon quadrupole at reionization also differs, and accordingly, the reionization bump is suppressed for the OCT mode. The oscillatory features also differ between the two tensor modes. Since the OCT metric perturbation is generated with a different phase, the peak positions in the tensor $EE$ and $BB$ spectra do not exactly coincide with those of the GW mode. These phase shifts are a characteristic feature of the tensor-OCT mode, and the $BB$ spectrum gives the clearest distinction between the two modes.

\begin{figure}[t]
\centering
\includegraphics[width=0.49\textwidth]{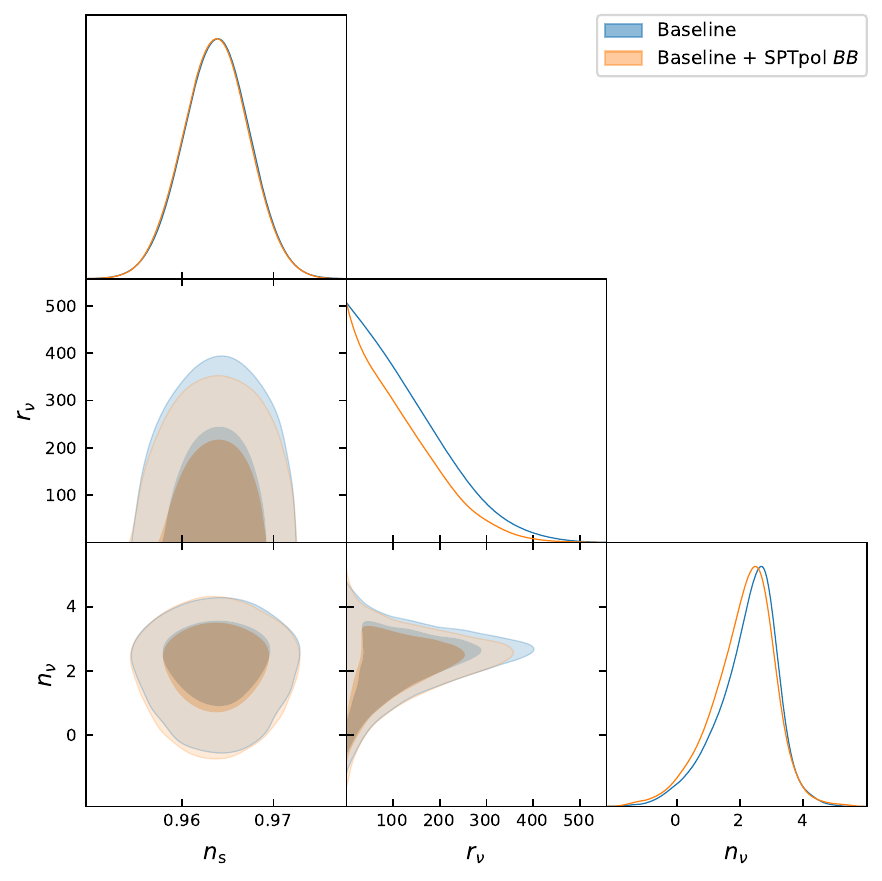}
\caption{Constraints on $(n_s,r_{\nu},n_{\nu})$ for the tensor-OCT mode. Blue and orange contours show the results for the baseline data and the baseline data combined with the four highest-multipole SPTpol $BB$ bandpowers, respectively. The dark and light contours represent the marginalized $68\%$ and $95\%$ credible regions for the stated priors, respectively.}
\label{fig:constraints}
\end{figure}

For the parameter constraints, we use \texttt{cobaya} \cite{2021JCAP...05..057T}. We consider only the OCT tensor mode and fix the standard GW amplitude to $r_{t}=0$. 
Our baseline likelihood consists of the DES five-year Type Ia supernova sample \cite{2024ApJ...973L..14D}, the SDSS-III BOSS DR12 consensus BAO likelihood \cite{2017MNRAS.470.2617A}, the Planck 2018 low-$\ell$ temperature and $E$-mode polarization likelihoods \cite{2020A&A...641A...5P}, the Planck PR4/NPIPE CamSpec high-$\ell$ $TTTEEE$ likelihood \cite{2022MNRAS.517.4620R}, the Planck 2018 lensing likelihood \cite{2020A&A...641A...8P}, and the BICEP/Keck 2018 likelihood \cite{2021PhRvL.127o1301A}. We additionally consider the baseline data combined with the foreground-marginalized SPTpol $BB$ likelihood \cite{2020PhRvD.101l2003S,2025OJAp....847459B}, retaining its four highest-multipole bandpowers.
We vary the six standard parameters of the flat-$\Lambda$CDM model together with $r_{\nu}$ and $n_{\nu}$, whose uniform priors are given by $r_{\nu} \in [0,5000]$ and $n_{\nu} \in [-3,10]$. We include lensing effects on all the theoretical CMB angular power spectra. We require the Gelman-Rubin convergence criterion $R-1<0.01$.

Figure~\ref{fig:constraints} shows the marginalized posterior distributions in the $(n_s,r_{\nu},n_{\nu})$ parameter space for the baseline data and the baseline data combined with SPTpol $BB$.
For the baseline data, power-law parametrization and priors described above, we obtain the marginalized $95\%$ Bayesian upper limit $r_{\nu} < 312$. The corresponding marginalized $95\%$ credible interval is $n_{\nu} = 2.2^{+1.6}_{-2.3}$, but it is not an independent measurement of the tilt in the absence of a detected OCT contribution. 
Including the selected SPTpol $BB$ bandpowers, we obtain the marginalized $95\%$ upper limit $r_{\nu}<283$ and the marginalized $95\%$ credible interval $n_{\nu}=2.1^{+1.7}_{-2.3}$. The additional high-$\ell$ $BB$ data thus tighten the upper limit on $r_{\nu}$ by approximately $10\%$, while leaving the marginalized interval for $n_{\nu}$ nearly unchanged.
At $r_{\nu} = 0$, the likelihood is independent of $n_{\nu}$. Marginalization over the nonnegative amplitude nevertheless weights tilts differently because the range of amplitudes compatible with the data depends on the spectral shape. The finite tilt interval thus reflects the likelihood together with the prior-weighted volume of allowed amplitudes, and it does not exclude any of the sampled tilts at arbitrarily small $r_{\nu}$. These numerical summaries are conditional on the adopted priors. A different amplitude prior, including a new flat prior imposed at another pivot, can change the marginalized tilt distribution.
The positive correlation between $r_{\nu}$ and $n_{\nu}$ mainly reflects the choice of pivot scale, especially in the B-mode spectrum.
The OCT contribution is strongly suppressed below $\ell \lesssim 200$, which roughly corresponds to the BK18 scales, as seen in Fig.~\ref{fig:spectra}, and the baseline observational constraint is therefore largely determined by the peak around this scale. By contrast, the adopted pivot scale, $0.05\,{\rm Mpc}^{-1}$, corresponds approximately to $\ell \sim 700$.

\begin{figure}[t]
\centering
\includegraphics[width=0.49\textwidth]{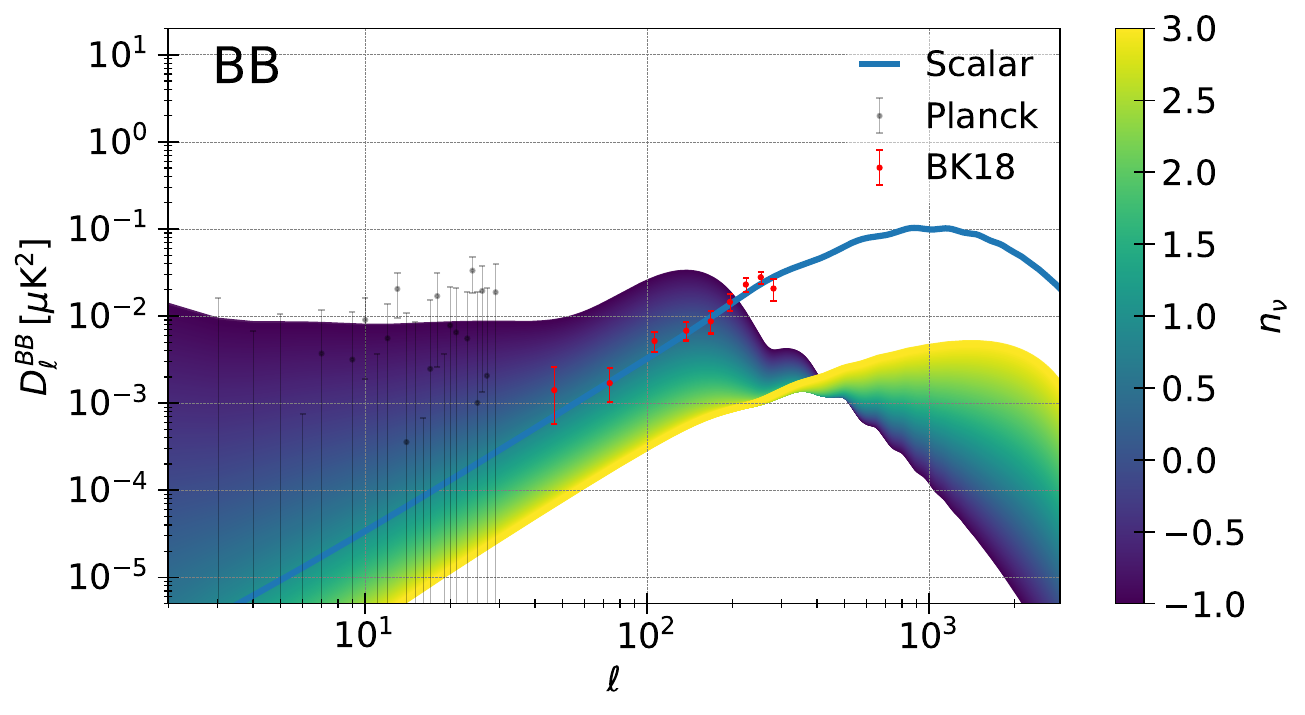}
\caption{CMB B-mode angular power spectrum for successively varying $n_{\nu}$ from $-1$ to $3$ but fixing $r_{\nu}=100$. The Planck and BICEP/Keck observational data (BK18), together with the lensing-induced scalar B-mode contribution, are also overplotted for comparison, as in Fig.~\ref{fig:spectra}.}
\label{fig:Dell_nnu}
\end{figure}

Figure~\ref{fig:Dell_nnu} shows the B-mode spectrum for successively varying $n_{\nu}$ with $r_{\nu}=100$. For a fixed amplitude at the pivot, increasing the blue tilt reduces the power around the lower wavenumbers relevant to the peak. A larger value of $r_{\nu}$ is consequently required to maintain a similar signal near $\ell \sim 200$. 
We note again that the numerically large value of $r_{\nu}$ does not directly mean a large metric perturbation and cannot be directly compared with the inflationary tensor-to-scalar ratio $r_{t}$. A blue power-law spectrum should also not be extrapolated to arbitrarily small scales. A physical generation model must specify the range of validity, or equivalently the ultraviolet cutoff, of the OCT spectrum.

\section{Summary}
\label{sec: summary}

In this paper, we investigated a regular tensor mode arising from the collisionless neutrinos. The anisotropic stress in the energy-momentum tensor is known to be a source of the tensor fluctuations, and the quadrupole of the collisionless neutrino distribution can be related to the anisotropic stress. However, a constant and uncompensated quadrupole cannot give a regular tensor mode. In contrast, although the octupole does not enter the Einstein equation directly, it first generates the quadrupole through the collisionless hierarchy, and the induced quadrupole, that is, anisotropic stress, sources the tensor metric perturbation. Consequently, both the quadrupole and the tensor perturbation start at first order in $k\eta$ and smoothly vanish in the super-horizon limit. Higher initial multipoles generate the metric only at higher orders in $k\eta$, and hence the octupole gives the leading moment of this class of regular tensor modes. We focused on this leading case in which the primordial amplitude is carried by the neutrino octupole, dubbed the tensor-OCT mode. We derived the corresponding initial conditions for the Einstein-Boltzmann system, implemented this mode in the public Boltzmann code, CAMB, and calculated its transfer functions and CMB angular power spectra.

The numerical solution of the tensor-OCT mode follows the expected hierarchy: the initial octupole sources the quadrupole, and the quadrupole then sources the metric perturbation. This hierarchical structure differs from the standard GW mode, and hence the induced metric perturbation has a different amplitude and phase from the standard GW mode. These differences clearly appear in the CMB spectra. In particular, the large-scale CMB anisotropies are suppressed both by the absence of a constant super-horizon metric perturbation and by the weak conversion from the octupole to the metric perturbation. The photon quadrupole generated at reionization is therefore smaller, leading to a suppressed reionization bump in the polarization spectra. The different phase of the metric evolution also shifts the oscillatory features in the CMB spectra. Therefore, the OCT contribution cannot be described by a simple rescaling of the standard GW mode, and the $BB$ spectrum provides the cleanest way to distinguish the two modes. Using the current observational data, we performed a Markov-chain Monte Carlo (MCMC) analysis of the primordial amplitude and spectral index of the tensor-OCT mode. 
The OCT contribution remains consistent with zero, with the
marginalized $95\%$ Bayesian upper limit $r_{\nu}<312$ for the baseline data and stated priors, tightening to $r_{\nu}<283$ when the selected SPTpol $BB$ bandpowers are included.
We also observed the positive correlation between $r_{\nu}$ and $n_{\nu}$, related to the choice of pivot scale of the primordial power spectrum.

The microscopic origin of the primordial neutrino octupole is still an open question. In the standard thermal history, neutrinos are collisional before decoupling, and their higher angular moments are driven toward zero. Therefore, the collisionless OCT solution should be regarded as an effective initial condition imposed at or after the beginning of free streaming. Possible origins may include nonstandard neutrino interactions or the decay of another component into collisionless particles. Constructing such a microscopic model is an important future direction. On the observational side, improved large-scale polarization measurements together with future small-scale $B$-mode observations will provide a useful test of the characteristic multipole dependence of the tensor-OCT mode based on our formalism.

% PTEP %%%%%%%%%%%%%%%%%%%%%%%%%%%
\ifPTEP
\section*{Acknowledgments}
This work was supported in part by the Japan Society for the Promotion of Science (JSPS) KAKENHI Grant Numbers JP24K17043 and 26H02044~(SS) and JP23H00108 and JP24K00627~(SY). SS also acknowledges support from DAIKO FOUNDATION and Hirose Foundation.
\fi
% PTEP %%%%%%%%%%%%%%%%%%%%%%%%%%%

% arxiv %%%%%%%%%%%%%%%%%%%%%%%%%%%
\ifarxiv
\begin{acknowledgments}
This work was supported in part by the Japan Society for the Promotion of Science (JSPS) KAKENHI Grant Numbers JP24K17043 and 26H02044~(SS) and JP23H00108 and JP24K00627~(SY). SS also acknowledges support from DAIKO FOUNDATION and Hirose Foundation.
\end{acknowledgments}
\fi
% arxiv %%%%%%%%%%%%%%%%%%%%%%%%%%%

% PTEP %%%%%%%%%%%%%%%%%%%%%%%%%%%
\ifPTEP
\bibliographystyle{ptephy}
\fi
% PTEP %%%%%%%%%%%%%%%%%%%%%%%%%%%
\bibliography{ref}
\end{document}